\documentclass[aip,jcp,reprint]{revtex4-1}
\usepackage{MnSymbol}
\usepackage{graphicx}
\usepackage{epstopdf}

\usepackage{dcolumn}
\usepackage{bm}

\graphicspath{C:/Users/duska/Disk Google/_MD_publikace/Comun_2017/Fig}

\draft 

\begin{document}


\title{Communication: Two-structure thermodynamics unifying all scenarios for water anomalies} 



\author{Michal Du\v{s}ka}
\affiliation{Institute of Thermomechanics of the CAS, v. v. i., Dolej\v{s}kova 1402/5, Prague 182 00, Czech Republic}
\affiliation{Institute for Physical Science and Technology, University of Maryland, College Park, Maryland 20742, USA}


\date{\today}

\begin{abstract}
Anomalous behavior of water in superooled region (namely decrease of density and sharp increase of response functions at atmospheric pressure) are manly associated with either existence of liquid-liquid criticality or re-entering vapor-liquid spinodal into positive pressure.  Despite different origin both scenarios introduce divergence of the response functions. We articulate in the communication that the criticality is behind water anomalies, water has normal vapor-liquid spinodal and that the re-entering feature was predicted because of curved density surface as a consequence of existence of critical point.  We can proof this by new two structure equation of state with criticality resulting from non-ideal mixing of two states and a background state being modeled by exactly the same equation of state which was deployed for predicting of vapor-liquid spinodal while formulating the re-entering scenario.
\end{abstract}


\maketitle 


Nature of water anomalies \citep{ZHELEZNYI1969, Rasmussen1973a, ANGELL1976, Kanno1979} remains unresolved mystery for centuries, hidden in experimentally inaccessible no-man’s-land beyond homogeneous nucleation limit of bulk water \cite{Kanno1975}. Numerous indicia we have gained from numerical experiments with atomistic models \citep{Russo2014, Singh2016, Moore209, Limmer2011, Limmer2013}, experiments with deeply suppercooled water confined in nonporous materials \citep{Liu2005, Zhang2011}, experimental study of microscopic configuration of bulk water \citep{Taschin2013, Huang2009}, and also from recent experiments with doubly metastabile water with respect to freezing and also evaporation \citep{Pallares2016}. Unfortunately unifying thermodynamic theory is style missing \citep{Gallo2016}. 

We have three candidates: Speede's hypothesis of re-entrant limit of stability of the liquid phase \cite{Speed1976}; Poole’s and his coworker's hypothesis of liquid-liquid critical point (LLCP) \citep{Poole1992} existing in no-man's-land with two special cases, critical point hidden in zero temperature the singularity free scenario \citep{Sastry1996} and critical point hidden beyond vapor-liquid spinodal the critical point free scenario \citep{Angell2008}; and the last one is quite recent based on extensive analyses of numerical experiments with atomistic models \citep{Limmer2015} authors associates large density fluctuations, thought to be a feature of coexistence of low density and high density water, with freezing like fluctuations.  

The first hypothesis was formulated after a decade of intensive exploring of water response functions in metastable region with respect to ice and their possible divergence beyond homogeneous nucleation limit. Speedy come with very elegant way to explore water vapor instability which indicated possible connection between these two divergences in a vapor-liquid spinodal re-entering to positive pressure.  This idea was backed only by lattice models \citep{Sastry1993, Borick1995} and very recently by numerical experiments with special "patchy" particles \citep{Rovigatti2017} but remained of the main argument mainly because a spinodal crossing its bimodal is impossible without another vapor liquid critical point \citep{Debenedetti2003}. 

Development of thermodynamic models suitable for studying the second scenario of water begins long before \citep{Strassler1965} it was postulated on bases of molecular dynamic study. Chemical reaction approach was used \citep{Moynihan1996, Ponyatovsky1998} to specifically address liquid-liquid water criticality or similarly formulated two-state model based on the concept of local symmetry \citep{Tanaka2000}. Two structure equation of state (TSEOS) was even adopted as international guideline for supercooled ordinary water by the International Association for the Properties of Water and Steam \citep{Holten2014}.

\

We will argue in this communication that in the first approximation all three scenarios can be unified in one simple approach and that heavy water instead of ordinary water is the right candidate to resolve nature of water anomalies. 

First step to the unified approach was work by Angell \citep{Angell1971}, he found two-state thermodynamics to be an approximate solution of "bond lattice". The idea of water as a mixture of two distinct species can be seen as counterintuitive, but the model (Gibbs free energy of water $G$ as a function of reduced temperature $ \hat{T} = T/T_{\mathrm{VLCP}}$   and pressure $ \hat{p} = p/p_{\mathrm{VLCP}}$; where VLCP indicates conditions at vapor liquid critical point)

\begin{eqnarray}
\frac{G}{k_{\mathrm{B}}\hat{T}}=&&
\frac{G^{\mathrm{A}}}{k_{\mathrm{B}}\hat{T}}+
x \frac{\left( G^{\mathrm{B}} - G^{\mathrm{A}} \right) }{k_{\mathrm{B}}\hat{T}}+
x \mathrm{ln}x \nonumber\\
&&
 + \left( 1 - x \right) \mathrm{ln}\left( 1 - x \right)  +
\frac{\omega}{k_{\mathrm{B}}\hat{T}} x \left( 1 - x \right)
\label{eq:1}
\end{eqnarray}
  	
addresses the main issue of broken hydrogen bonds very straightforwardly as non-ideal mixture of tetrahedral water $ G^{\mathrm{B}} $ with fraction of its molecules $x$ and water with broken hydrogen bonds $ G^{\mathrm{A}} $  normal liquid (where $ k_{\mathrm{B}} $ is Boltzmann’s constant, and  $ \omega = \omega_{0}\left( 1 + \omega_{1}\hat{p} + \omega_{2}\hat{T} \right)  $  is the cooperativity controlling preference of the species for its neighbors and if the two spices do not like each other separation of the two spices begun in a LLCP). Coexistence of very different structures is not unusual; for example in water steam there is always, thanks to fluctuations, some "concentration" of water like clusters increasing their size and numbers while approaching saturation and even more with supper saturation. What is special about liquid water is immunity of "tetrahedral water", or if we want "freezing like fluctuations" \citep{Limmer2011, Limmer2013}, towards irreversible freezing. This unusual property results in strong presence of such a structure in liquid water \citep{Zhang2011, Huang2009}. The best prove of this resistance is existence of low density amorphous ice the tetrahedral like structure in amorphous form \citep{Bellissent1995}. "Tetrahedral water" can be as a small cluster in a steam result of a mere fluctuation \citep{Limmer2011, Limmer2013, Limmer2015} but this is not contradiction to possibility of very complicated interplay between these distinct structures of water possibly resulting in a phase separation ended in critical point faraway from equilibrium condition observed in many numerical experiments with atomistic models \citep{Russo2014, Singh2016, Moore209, Limmer2011, Limmer2013}.  Especially when times required for reorganizing atomic arrangements in the liquid (structural relaxation time) are significantly faster than times required for nucleating and growing a crystal, even faraway from equilibrium \citep{Limmer2011}. Therefore we will apply kinetically persistent chemical equilibrium to find "equilibrium mixture" of the two structures from 

\begin{eqnarray}
\frac{\partial G/k_{\mathrm{B}}\hat{T}}{\partial x}=0.
\label{eq:2}
\end{eqnarray}

Based on this consideration we will use the two-state thermodynamics as a first order approximation of interplay of water structures regardless of their origin. 

\begin{figure*}
  \renewcommand{\arraystretch}{0}
  \begin{tabular}{@{}cccc@{}}
    \includegraphics[scale=1]{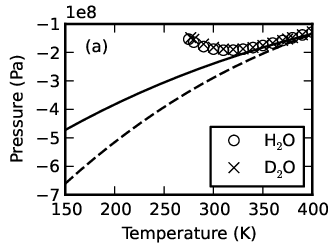} &
    \includegraphics[scale=1]{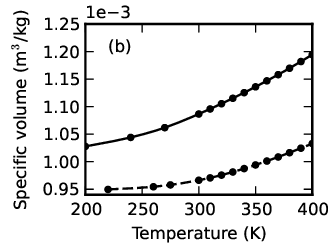} &
    \includegraphics[scale=1]{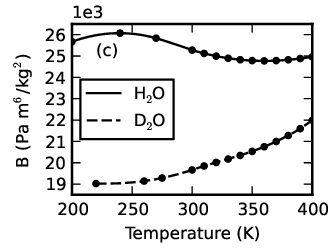} &
  \end{tabular}
  \caption{\label{fig:wide1} For both substances: (a) Spinodal pressure calculated directly from Speede's equation (points), and spinodal pressure of state A (curves) represented in model by polynomial function of temperature; (b) volume at spinodal; (c) and second derivative of pressure with respect to density at constant temperature at spinodal both represented by cubic spline of the pints $\bullet$. }
\end{figure*}

The last step to find unified approach is connecting liquid-liquid criticality to Speed's hypothesis. In this effort the simplification by two-state approach is very useful. We assume that state A as the "normal liquid" can be described by Speedy's equation of state 

\begin{eqnarray}
\hat{V}_{\mathrm{A}} = 
\left( \frac{\partial G_{\mathrm{A}}}{\partial \hat{p}} \right)_{T} =
\frac{\hat{V}_{\mathrm{SP}} \left( \hat{T}\right) \sqrt{B \left( \hat{T} \right)  } }{\sqrt{1-\hat{p}/\hat{p}_{\mathrm{SP}} \left( \hat{T}\right)} + \sqrt{B \left( \hat{T} \right)} }
\label{eq:3},
\end{eqnarray}

and that all anomalous behavior even the re-entrant spinodal can be fully explained by criticality. All the three temperature dependent functions from equation above are presented in Fig.~\ref{fig:wide1}. The pressure at spinodal $ \hat{p}_{\mathrm{SP}} \left( \hat{T}\right) $   smoothly connects to Speed's limit of stability at higher temperatures suggesting a vanishing of tetrahedral water. Despite proximity of the Speed's limit of stability for both liquids the state A of heavy water (we shell point out that limit of stability of our simple two state equation is very close to stability of the state A) seems to be far more stable with respect to stretching then ordinary water according to our model. Resistance towards stretching could be explained by stronger hydrogen bonds in heavy water \cite{Soper2008}. The volume at spinodal $ \hat{V}_{\mathrm{SP}} \left( \hat{T}\right) $  is monotonic function with gradually decreasing decline at supercooled region. The second derivative of pressure with respect to density at constant temperature at spinodal $ B \left( \hat{T} \right) $ is little bit more complicated function, based on theory it has to be zero at VLCP so it has to turn sharply down in higher temperature. 

The last part of two state model to be defined is difference between Gibbs free energy of state B and A; which is besides of specifying thermodynamic property of pure state B also responsible for position of liquid-liquid transition line (LLTL) and "Widom" line (WL) where $ x = 0.5 $ for the symmetric chemical reaction (coming from Eqs.~\ref{eq:2})

\begin{eqnarray}
&&\frac{\left( G^{\mathrm{B}} - G^{\mathrm{A}} \right) }{k_{\mathrm{B}}\hat{T}} = \\
&&\lambda \left( \frac{1}{\hat{T}} +
 a_{0} \hat{p} +
 a_{1} \frac{\hat{p}}{\hat{T}} +
 a_{2} +
 a_{3} \hat{T} 
 + a_{4} \frac{\hat{p}^{2}}{\hat{T}} +
 a_{5} \frac{\hat{p}^{3}}{\hat{T}} +
 a_{6} \hat{p}\hat{T} \right) \nonumber
\label{eq:4}.
\end{eqnarray}

\begin{figure*}
  \renewcommand{\arraystretch}{0}
  \begin{tabular}{@{}cccc@{}}
    \includegraphics[scale=1]{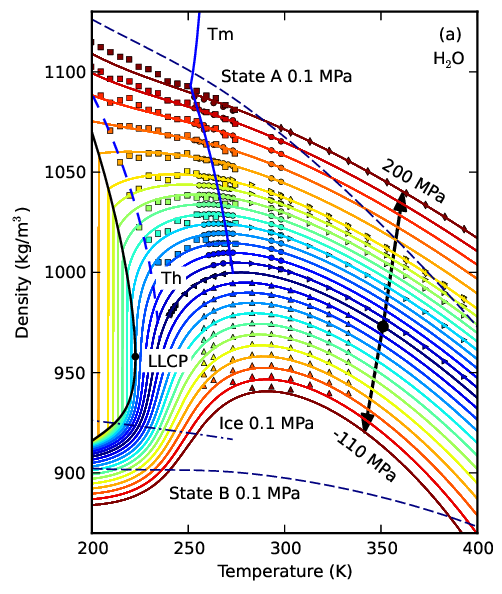} &
    \includegraphics[scale=1]{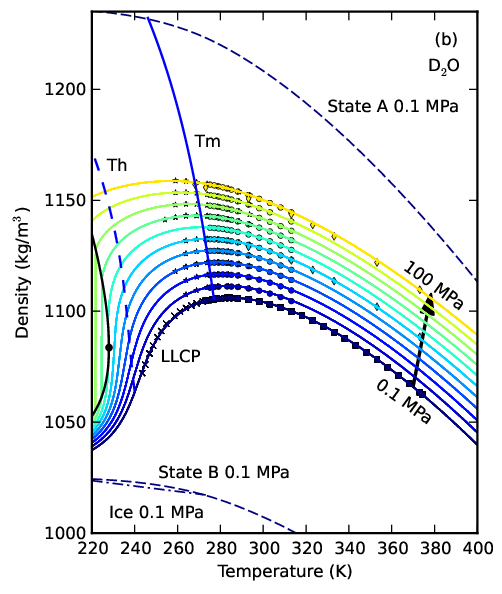} &
  \end{tabular}
  \caption{\label{fig:wide2} Density of the model of (a) ordinary water:  for pressures at 10 MPa intervals between -110 MPa and 0.1 MPa and  100 MPa, and then at 20 MPa intervals up to 200 MPa is compared to the experimental data of Mishima (2010) \citep{Mishima2010} $\blacksquare$, Hare end Sorensen (1985) \citep{Hare1987} $\blacktriangleleft$, Sotani \textit{et al.} (2000) \citep{Sotani2000} $\bullet$, Kell and Whalley (1975) \citep{Kell1975}  $\blacktriangleright$, Grindley and Lind (1971) \citep{Grindley1971} $\blacklozenge$, and Pallares \textit{et al.} (2016) \citep{Pallares2016} $ \blacktriangle $. Density of the model of  (b) heavy water: for pressures at 10 MPa intervals between 0.1 MPa and  100 MPa, is compared to the data of Rasmussen and MacKenzie (1973) \citep{Rasmussen1973} $\times$, Emmet and Millero (1975) \citep{Emmet1975} $\bullet$, Kell (1967) \citep{Kell1967} $\blacksquare$, Bridgman (1935) \citep{Bridgman1935} $\blacklozenge$ and Du\v{s}ka \textit{et al.} (2017) \citep{Duska2017} $\bigstar$ (pressure of the shown experimental data deviate less than 3\% from the model). Solid blue line with Tm is the density of water at melting point, dashed blue line with Th is the density at homogeneous nucleation limit, density of state A and B at atmospheric pressure are shown as dashed lines, density of ice at atmospheric pressure as dot and dashed line and liquid-liquid transition line ended at critical point is a solid black line with circle labeled with LLCP.}
\end{figure*}

All parameters of two state model and state A (given in Supplement) were optimized simultaneously with help of genetic optimization to fit only density data at positive pressure, because of special form of equation of state for state A, see Eqs.~\ref{eq:3}. This equation can be integrated to get Gibbs free energy of state A; but it will be implemented in the future.  Model's representation of experimental data is presented in Fig.~\ref{fig:wide2}. For both substances there is very good agreement with the date even in negative pressure but model for ordinary water has difficulty to match density of pure state B with density of ice (both tetrahedral structures) as well as density of very stretched water in low temperatures. In the case of ordinary water asymmetry in chemical reaction has to be introduce into our TSEOS.

\begin{figure*}
  \renewcommand{\arraystretch}{0}
  \begin{tabular}{@{}cccc@{}}
    \includegraphics[scale=1]{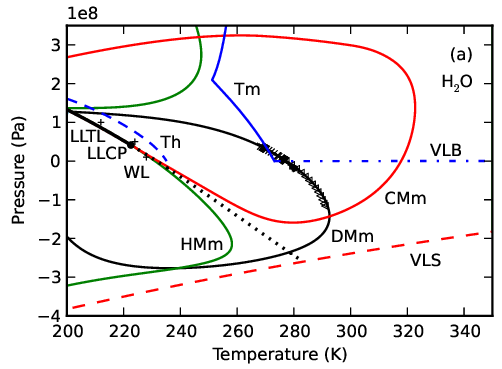} &
    \includegraphics[scale=1]{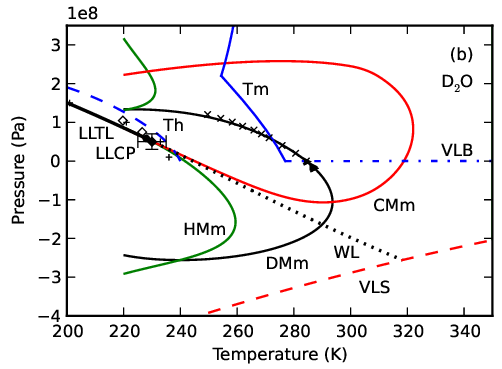} &
  \end{tabular}
  \caption{\label{fig:wide3} Phase diagram for ordinary water (a) and heavy water (b) with density extrema locus (DMm) as a black curve, compressibility extrema locus (CMm) as a  solid red curve and isobaric heat capacity extrema locus (HMm) as a green curve.  Where the WL is Widom line (dotted black line) crossing homogeneous nucleation line (Th) at positive pressure in case of heavy water, the VLB is vapor-liquid bimodal, the VLS is vapor-liquid spinodal and the Tm is melting line. For ordinary water DMm is compared to experimental data of Caldewll (1978)\citep{Caldwell1978} $ \times $ , Henderson and Speedy (1987) \citep{Henderson1987} $\blacktriangleright$ and Pallares \textit{et al.} (2016) \citep{Pallares2016} $ \lhd $; DMm of heavy water is compared to data of Kanno and Angell (1980) \citep{Kanno1980} $\times$ and Henderson and Speedy (1987) \citep{Henderson1987} $\blacktriangleright$. A prediction by Mishima (2000) \citep{Mishima2000} of a liquid-liquid critical point (LLCP) for heavy water is marked as $\blacklozenge$ and a liquid-liquid transition line (LLTL) as $\lozenge$. LLTL of model is thick black line and LLCP is marked as $\bullet$. For both substances projection of mechanical stability limit $+$ by Kanno and Angell (1979) \citep{Kanno1979} is also presented.}
\end{figure*}

It can be seen from Fig.~\ref{fig:wide2} that density of heavy water behaves very differently from ordinary water in supercooled region close to atmospheric pressure. There is a much sharper decline in experimentally acceseble region that even an inflection point could be crossed indicating proximity of Widom line. This is apparent form phase diagrams in Fig.~\ref{fig:wide3}; not only Widom line is more pronounced at higher temperatures with respect of homogeneous nucleation limit, crossing it at positive pressure (22 MPa), but also position of maximum of density with respect to melting line. We see this as a consequences of stronger tetrahedral structures in heavy water \cite{Soper2008}. According to our model both lines of maximal compressibility and isobaric heat capacity of heavy water are accessible for balk experiments. It means that if we are right existence of liquid-liquid critical point at non-accessible region for balk experiments could be proven by presence of Widom line in metastabile region of heavy water.

\

Equation of state described in the communication successful explanations nature behind re-entering spinodal hypotheses as a consequence of liquid-liquid critical point existence, therefore it unifies all scenarios to explain water anomalies.

\

We are grateful for collaboration with J. V. Sengers and for his suggestions to improve the article. The research of M.D. was supported by the International Association for the Properties of Water and Steam.

%

\bibliography{Com2017}

\end{document}